\documentclass{article}
\pdfoutput=1

\PassOptionsToPackage{numbers, compress}{natbib}

     \usepackage[preprint]{neurips_2019}

\usepackage[utf8]{inputenc} 
\usepackage[T1]{fontenc}    
\usepackage{hyperref}       
\usepackage{url}            
\usepackage{booktabs}       
\usepackage{amsfonts}       
\usepackage{amssymb}
\usepackage{amsmath,bm}
\usepackage{nicefrac}       
\usepackage{microtype}      
\usepackage[pdftex]{graphicx}
\usepackage{natbib}

\title{Residual Entropy}

\author{%
  Barnaby~Rowe \\
  Senior Data Scientist\\
  Fidelity International\\
  4 Cannon Street\\
  London EC4M 5AB\\
  United Kingdom\\
  \texttt{barnaby.t.p.rowe@gmail.com};\\
  \texttt{barney.rowe@fil.com} %
}

\begin{document}

\maketitle

\begin{abstract}
  We describe an approach to improving model fitting and model generalization that considers the
  entropy of distributions of modelling residuals.  We use simple simulations to demonstrate the
  observational signatures of overfitting on ordered sequences of modelling residuals, via the
  autocorrelation and power spectral density.
  These results motivate the conclusion that, as commonly applied, the least squares method
  assumes too much when it assumes that residuals are uncorrelated for all
  possible models or values of the model parameters.  We relax these too-stringent assumptions in
  favour of imposing an entropy prior on the (unknown, model-dependent, but potentially
  marginalizable) distribution function for residuals. 
  We recommend a simple extension to the Mean Squared Error loss function
  that approximately incorporates this prior and can be used immediately for modelling applications
  where meaningfully-ordered sequences of observations or training data can be defined.
\end{abstract}

\section{Introduction}\label{sec:intro}
The method of least squares is the dominant approach in regression problems. 
The Mean Squared Error
(MSE) is frequently therefore the \emph{default} loss function in many practical applications:
and is commonly used for regression in fields as diverse as machine learning, statistics,
physical sciences, decision theory, econometrics, and finance.  It is criticised for over-use and
a lack of robustness \citep[e.g.][]{berger85}, but has yet
to be displaced.
In this paper we will level
another criticism at the MSE and other similarly-constructed loss functions: they leave out
important information as a result of their too-strict assumptions about the uncorrelatedness of
model residuals.

Statistical analysis of correlated residuals has a long history in econometrics and time
series analysis, that began with tests of
the presence of autocorrelation at lag $l=1$ in residuals from least squares regression
\citep{vonneumann41,durbinwatson50,durbinwatson51,durbinwatson71}.
Tests were later developed for the presence of autocorrelation at any lag
\citep{boxpierce70,ljungbox78}.  The question became acute as econometrics developed,
as autocorrelations in residuals for autoregressive time series models
(where some of the regressors are lagged dependent variables) leads in general to
non-consistency and bias.
Sophisticated tests have therefore been developed to help validate such models
\citep{breusch78,godfrey78,proia18} and make them robust to further
confounding real-world factors such
as conditional heteroskedasticity \citep[e.g.][]{godfreytremayne05}.

While prior work in this area has to been to devise tests of autocorrelation that can be applied
\emph{after} best-fitting regression models have been determined, e.g.\ by least squares,
we propose that the loss function itself could be extended to penalize models
showing autocorrelated residuals \emph{during} the process of fitting.  Such correlations can
either be created by the presence of correlated errors in observations of the independent variable,
or by the application of the wrong model.
In the first Section we illustrate this for the important category of overfitting (or
over-specified) models by performing simple simulations of model fitting in one dimension.

In later Sections we make a simple entropy argument for an extension to the MSE loss function
that is readily and efficiently calculated, at least in one dimension or wherever residuals can
be meaningfully ordered (e.g. time series forecasting).  We illustrate from our simulations how this
modified loss function will effectively penalize overfitting models,
potentially providing natural regularization, better generalization and enhanced
robustness to outliers.

\section{Simulations of Overfitting in One Dimension}\label{sec:sims}
Consider a sample $Y = \left( y_0, y_1,\ldots,y_{N-1} \right)$ of $N$ data values for our dependent
variable of interest $y$, situated at corresponding locations
$\left(x_0, \ldots,x_{N-1} \right)$ in the indpendent variable.
We will construct a regression model for estimates of $y$ which we will denote
$\hat{y}(x | \bm{\theta})$ for a vector of model parameters $\bm{\theta}$.
The residuals $R$ between the model and the data are defined as $R = (r_0,\ldots, r_{N-1})$
where $r_n = y_n - \hat{y}(x_n)$ for each element $n=0,\ldots,{N-1}$ of the sample.

To explore the effects of overfitting, we will draw $Y$ as i.i.d.\ standard normal variables, the
distribution of residuals you would expect from a good fit to the data,
\emph{and then attempt to fit these values}.
The aim is to simulate what happens when a model is given unnecessary additional freedom.

\subsection{A Fourier Series Model}\label{sec:sinsims}
\begin{figure}
  \centering
  \includegraphics[width=0.49\textwidth]{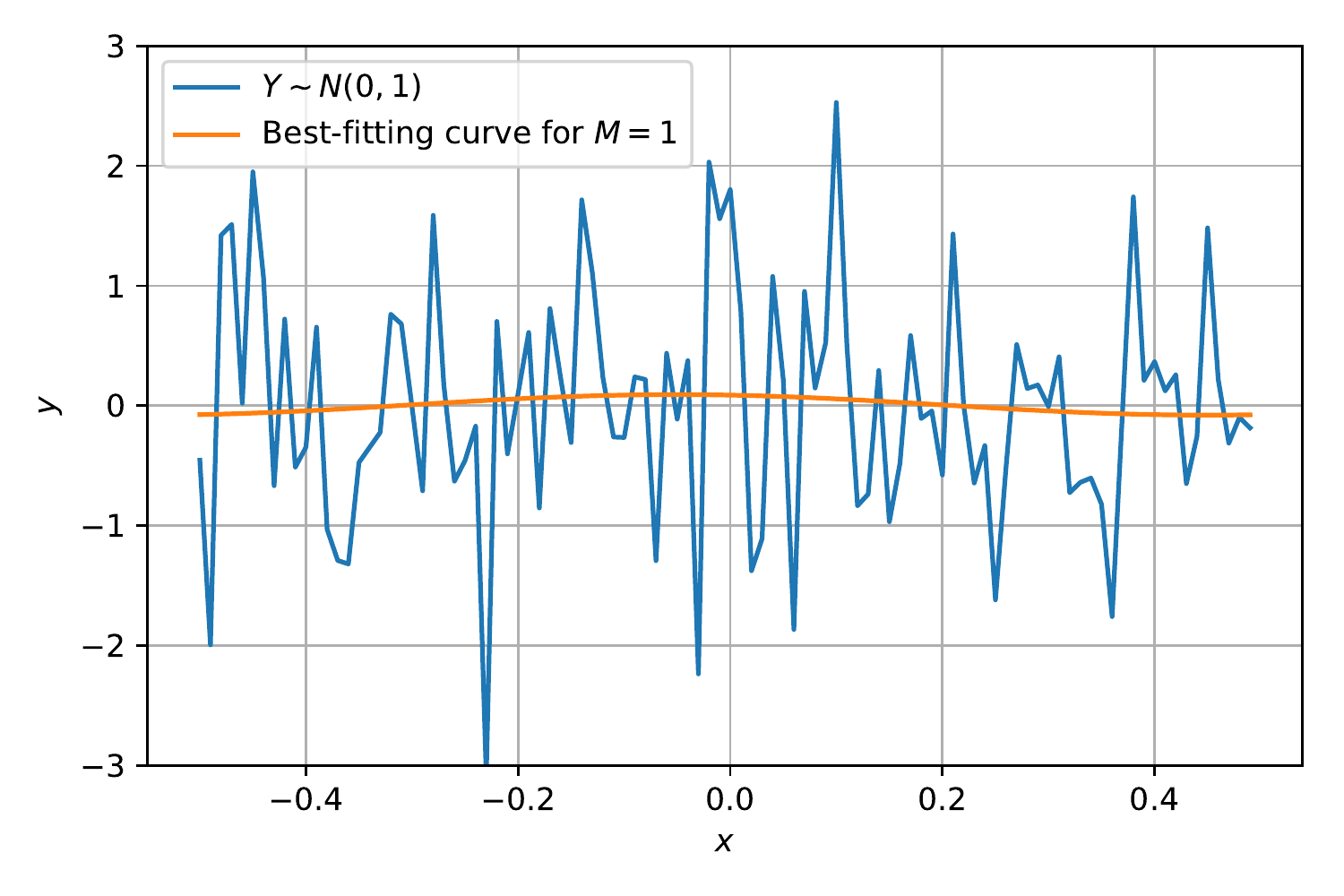}
  \includegraphics[width=0.49\textwidth]{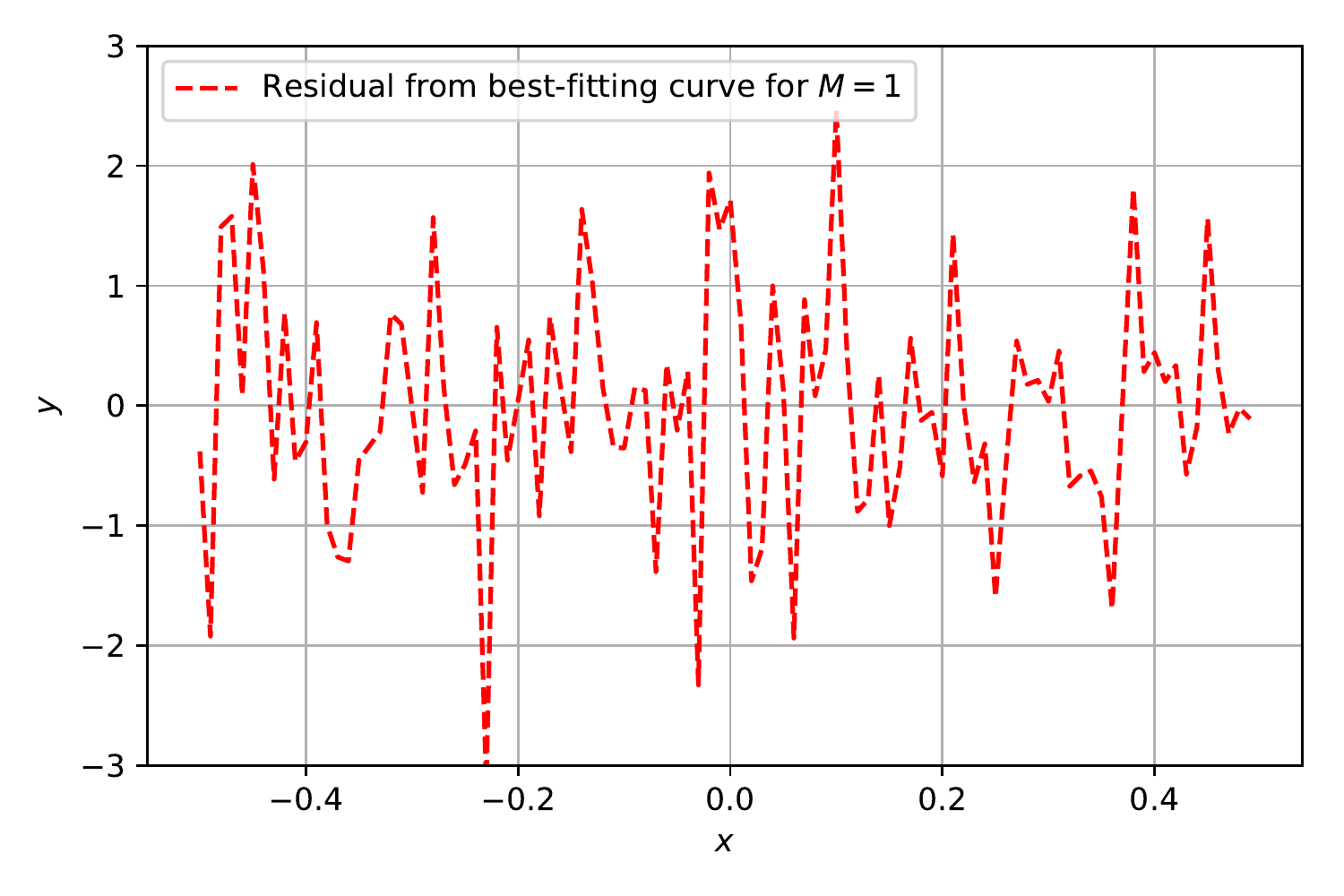}
  \includegraphics[width=0.49\textwidth]{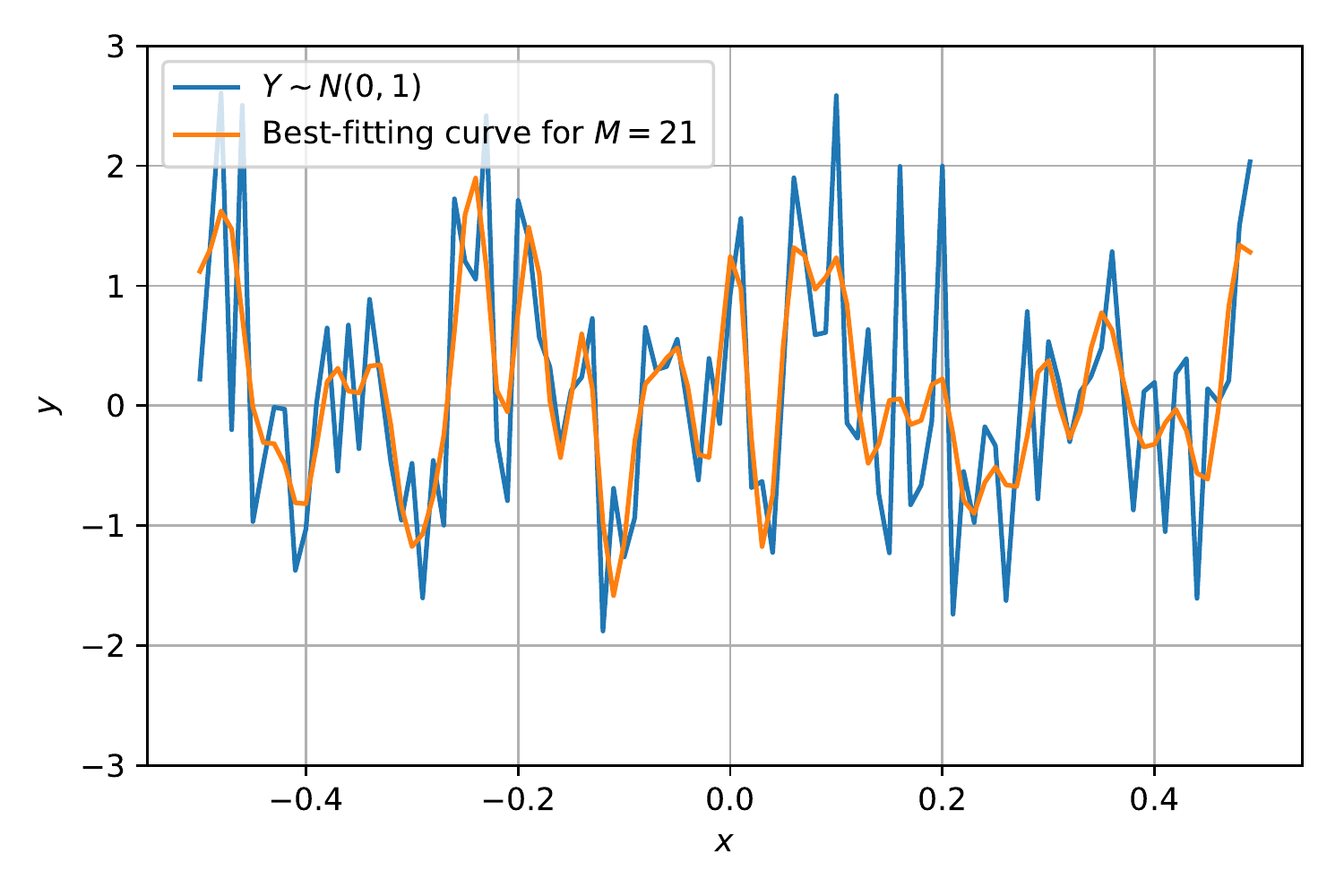}
  \includegraphics[width=0.49\textwidth]{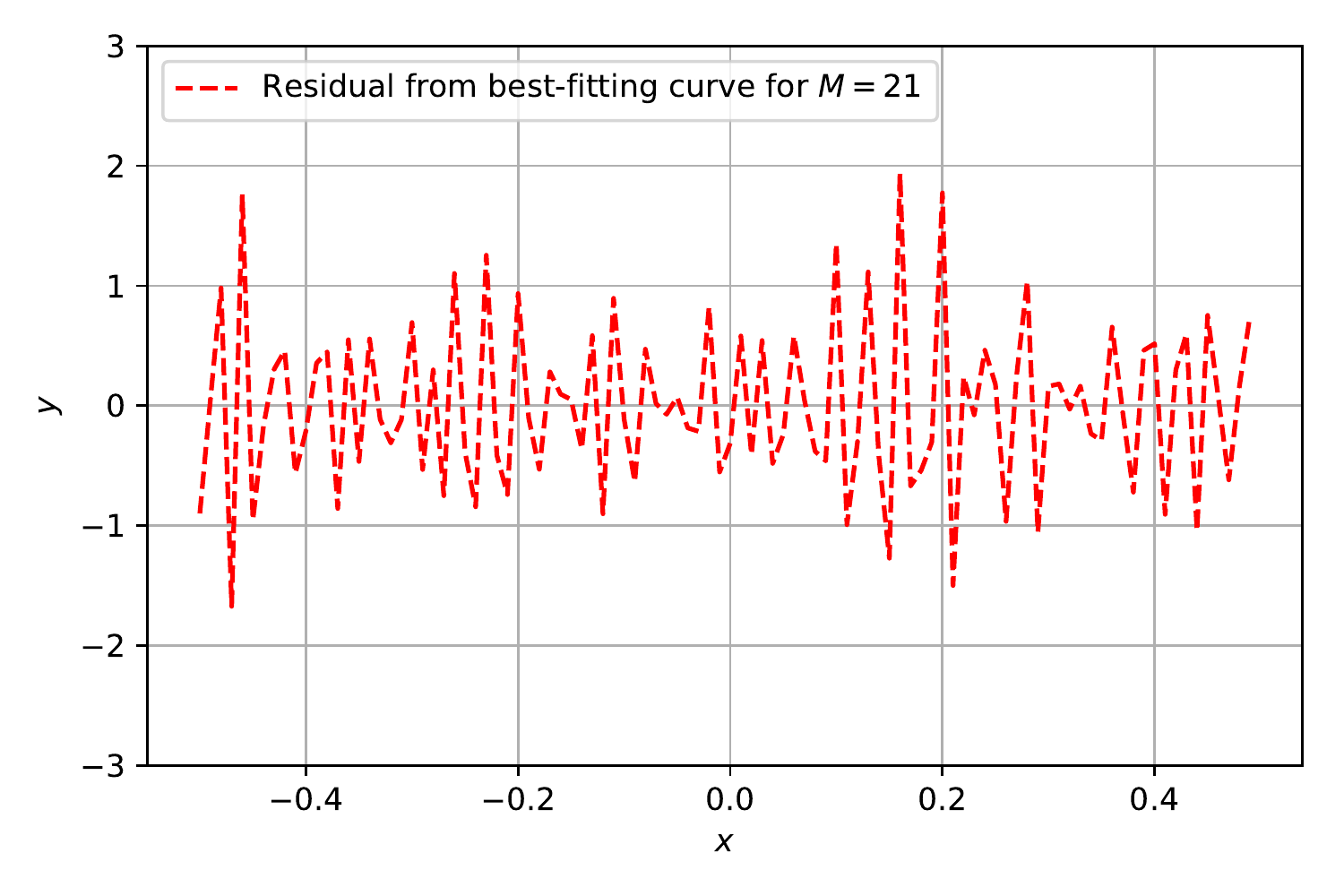}
  \includegraphics[width=0.49\textwidth]{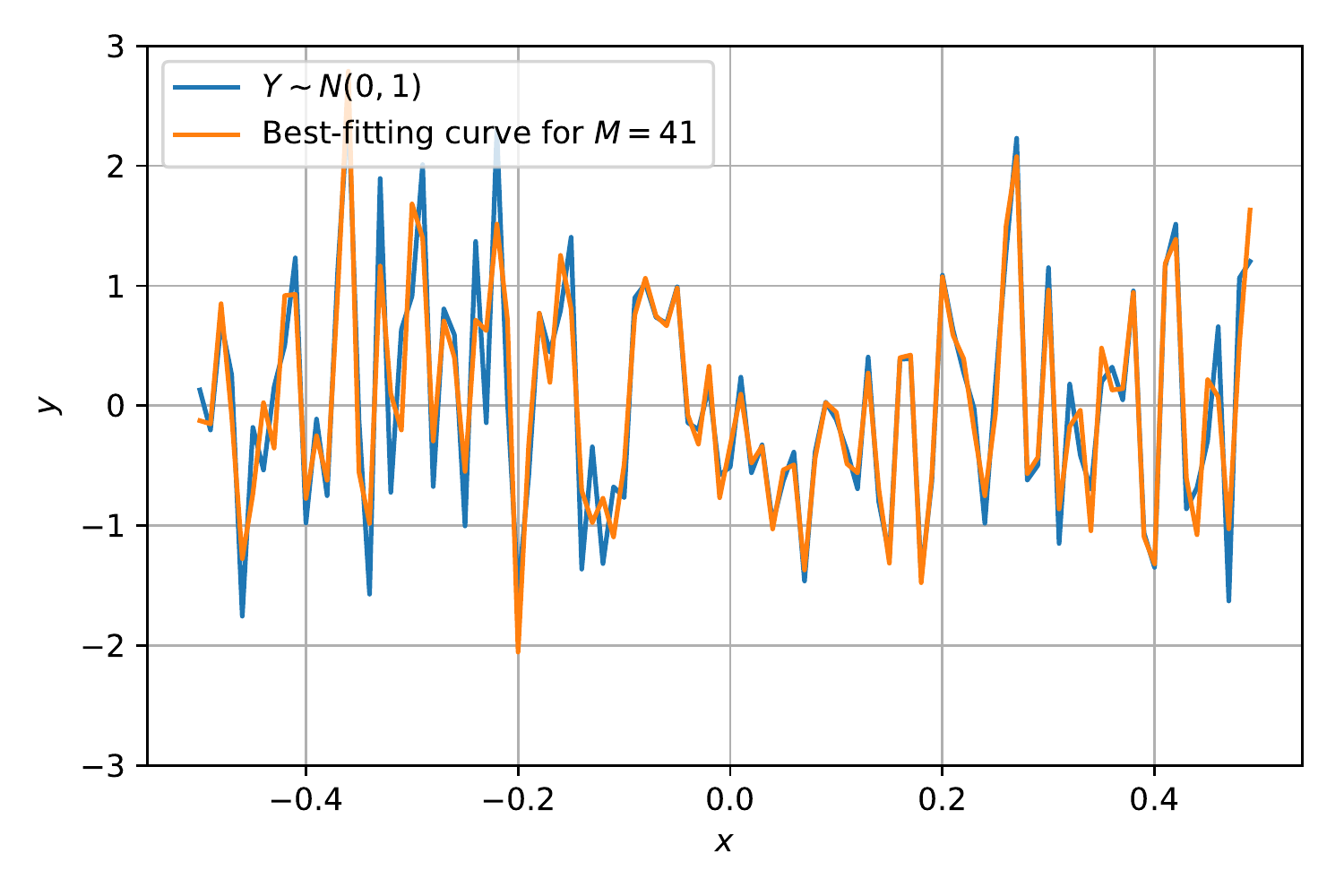}
  \includegraphics[width=0.49\textwidth]{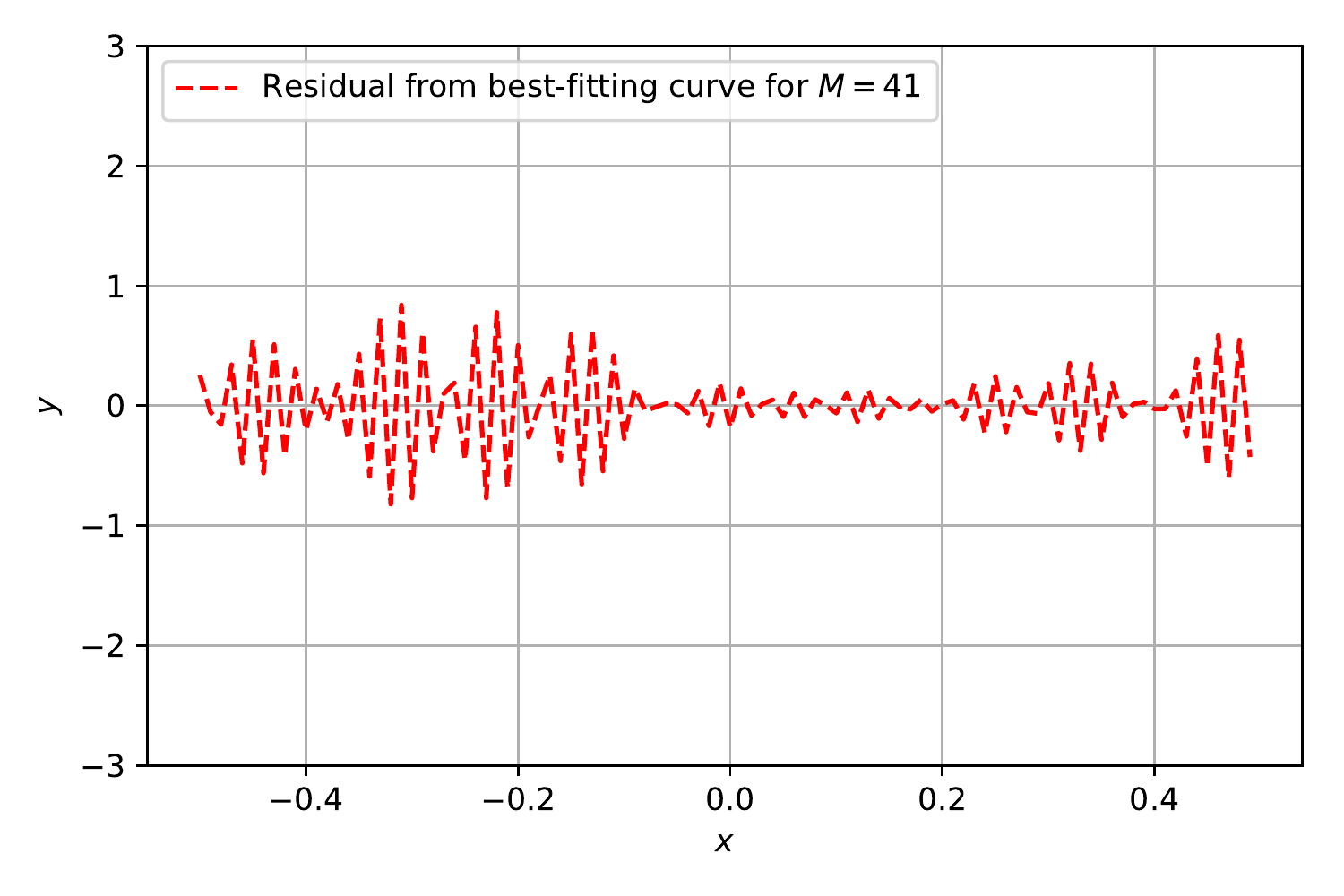}
  \caption{Three examples of randomly-drawn samples $Y$ and the least-squares best-fitting model
  (left column), and resulting residuals (right column), for model orders $M=1$ (top row),
  $M=21$ (middle row), and $M=41$ (bottom row) of the Fourier series model of equation
  \eqref{eq:sinusoids}.  The residuals reduce in
  amplitude but also appear ``less random'' as the model order increases.\label{fig:res}}
\end{figure}
We simulate a simple case of $N=100$ observations for our randomly-drawn sample $Y$,
with corresponding evenly-spaced locations $x_n = -0.5 + n/N$ for $n=0,\ldots,{N-1}$.
We define the model of order $M$ as a truncated Fourier Series:
\begin{equation}\label{eq:sinusoids}
\hat{y}_{\mathrm{s}, M} \left(x | a_0, a_1, \ldots, a_{M-1}, b_1, \ldots, b_{M-1} \right) =
\frac{1}{2} a_0 + \sum_{m=1}^{M-1} a_m \cos{(2 \pi m x)}  + b_m \sin{(2 \pi m x)} .
\end{equation}
Figure~\ref{fig:res} shows three examples of least-squares fitting
$\hat{y}_{\mathrm{s}, M}(x)$ to different randomly-generated samples $Y$, and the resulting
residuals.
We can see that as the overfitting order $M$ increases to 41 there is observable structure in the
sequence of residuals $R$.

The aim of this paper is to show that there is valuable information about modelling in the way that
residual values are distributed with respect to one another in the domain of input variables.
The tools of correlation and spectral analysis are the primary aids for assessing these properties.
We define the (circular) residual autocorrelation function at lag $l$ as
\begin{equation}
\rho_{rr}(l) \triangleq \left. \sum_{n=0}^{N-1} r_n ( r_N )^*_{n-l} \middle/ \:
\left( \sum_{n=0}^{N-1}r_n r^*_n \right) \right. ,
\label{eq:rho}
\end{equation}
where $r^*_n$ denotes the complex conjugate of $r_n$ (we will assume real-valued $r_n$ hereafter so
that $r_n r^*_n = r^2_n$),
and where $r_N$ denotes an element of
the sequence $R$ extended by periodic summation to an infinitely long, repeating sequence of
period $N$ so that for $p \in \mathbb{Z}$, we have $(r_N)_{p} = r_{(p \! \! \! \mod \! N)}$. 

The term in the denominator of \eqref{eq:rho}, commonly referred to as the Residual Sum of Squares
(RSS) or Sum of Squared Errors of prediction (SSE), normalizes the function so that
$\rho_{rr}(l=0) \triangleq 1$. At
lags $l \ne 0$ statistically independent residuals have $E(\rho_{rr}) = 0$.  For purely real
residuals \eqref{eq:rho} is an even symmetric function of $|l|$, and is Hermitian in general.

This definition of the residual autocorrelation function is convenient because it can be related
simply to the Discrete Fourier Transform (DFT) of the sequence of residuals
$\tilde{R} = \left( \tilde{r}_0, \ldots, \tilde{r}_k, \ldots, \tilde{r}_{N-1} \right)$,
where
\begin{equation}
\tilde{r}_k = \sum_{n=0}^{N-1} r_n \, e^{-\frac{i 2 \pi}{N} kn}
\end{equation}
and the inverse transform is defined as
\begin{equation}
r_n = \frac{1}{N} \sum_{k=0}^{N-1} \tilde{r}_k \, e^{\frac{i 2 \pi}{N} kn}.
\end{equation}
We will often shorthand the DFT operations on a full sequence using the symbol $\mathcal{F}$,
e.g.\ $\tilde{R} = \mathcal{F}\{R\}$ and $R = \mathcal{F}^{-1}\{\tilde{R}\}$.  Using the circular
form of the Wiener-Khinchin theorem for DFTs we may then write
$\rho_{rr}(l) = \mathcal{F}^{-1}\{ \tilde{R} \, \tilde{R}^* \} /
\left( \sum_{n=0}^{N-1} r^2_n \right)$, which can be calculated with great computational
efficiency thanks to the Fast Fourier Transform (FFT) algorithm \citep{cooleytukey65}.

\begin{figure}
  \centering
  \includegraphics[width=0.8\textwidth]{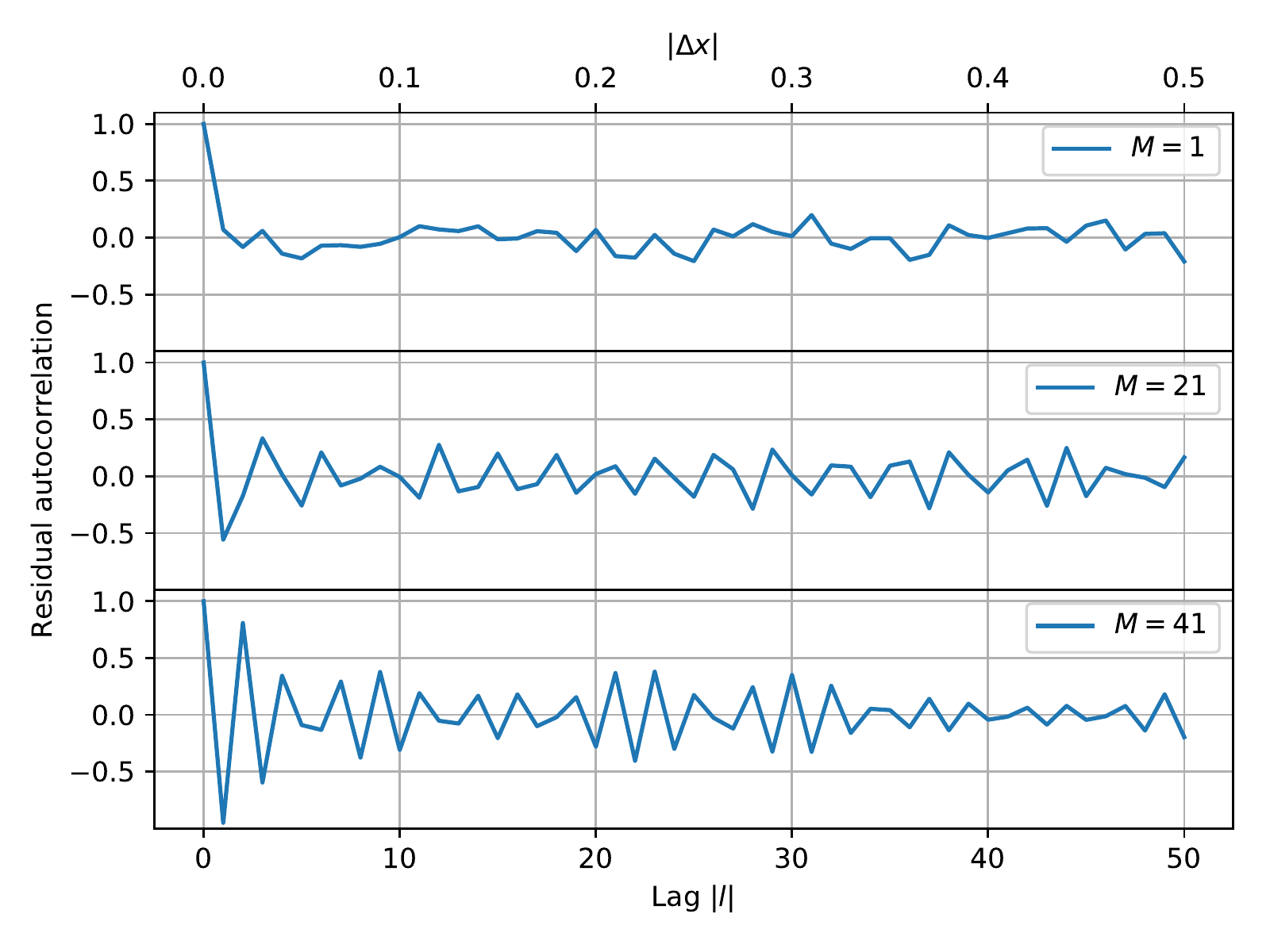}
  \caption{Residual autocorrelation function values for the randomly-drawn samples and 
  least-squares best-fitting models of Figure~\ref{fig:res}.\label{fig:rescor}}
\end{figure}
\begin{figure}
  \centering
  \includegraphics[width=0.49\textwidth]{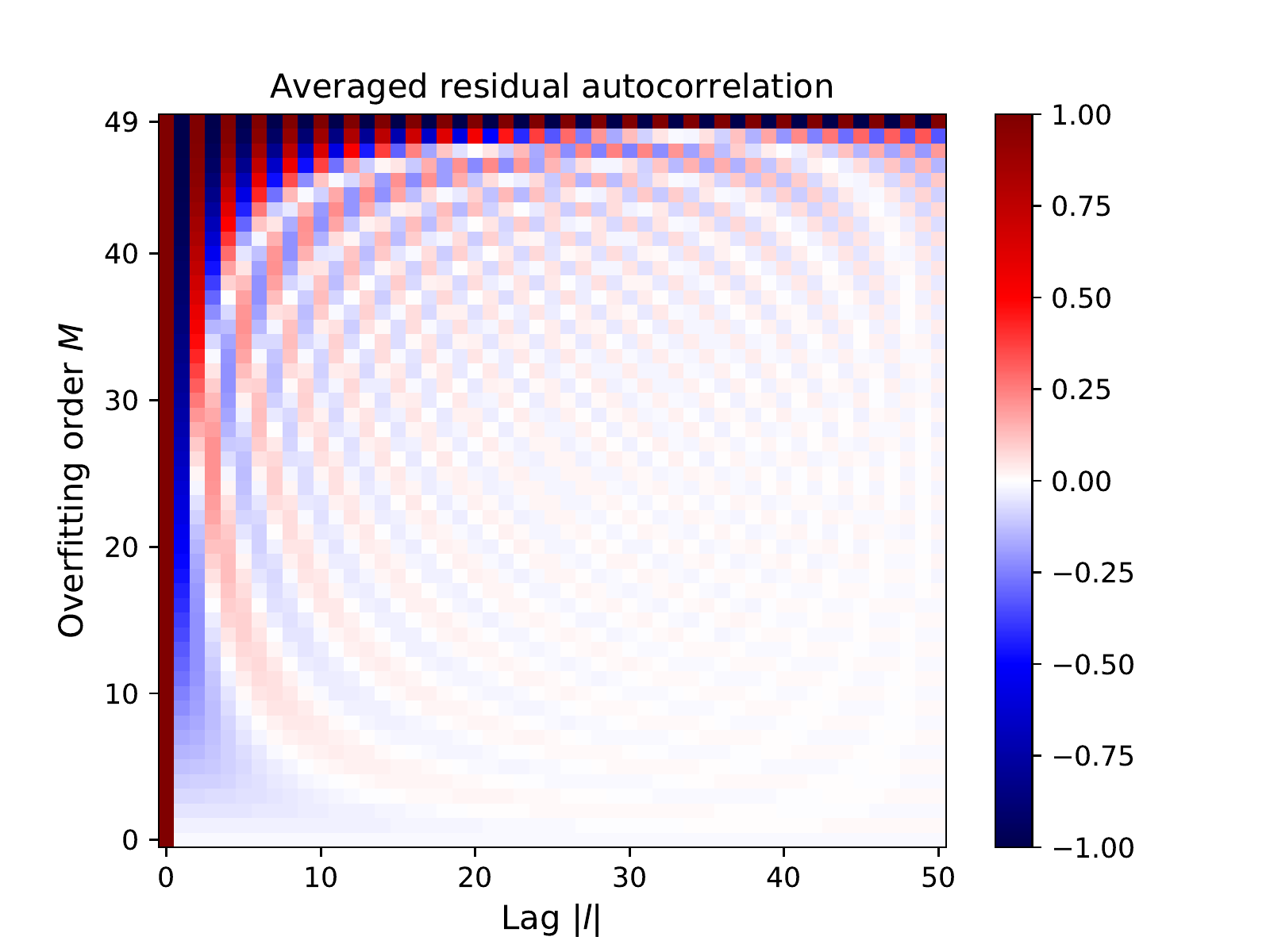}
  \includegraphics[width=0.49\textwidth]{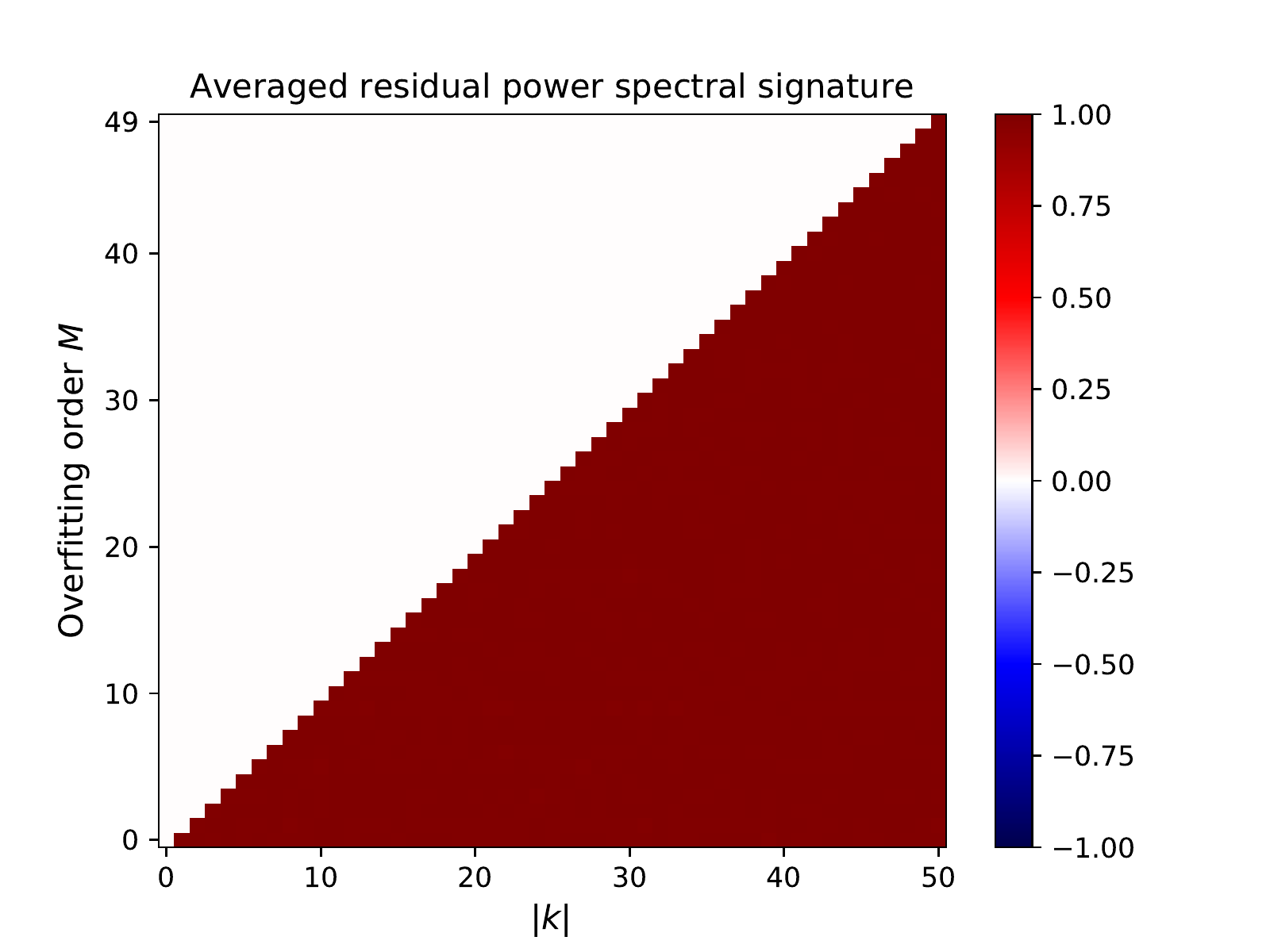}
  \caption{a) Averaged residual autocorrelation function values as a function of Fourier Series
  model order $M$,
  calculated over $10^5$ realizations of $Y$ per order; b) Averaged residual power spectral
  signature for the same realizations and model.\label{fig:scorall}}
\end{figure}

Figure~\ref{fig:rescor} shows $\rho_{rr}(l)$ for the
model fits of Figure~\ref{fig:res}.
Significant negative correlation between neighbouring residuals is seen
at lag $|l| = 1$ for the $M=21$ and $M=41$ models. The $M=41$ model shows the strongest
departures from $\rho_{rr} \simeq 0$ at lags $l \ne 0$, with an oscillating character.
Figure~\ref{fig:scorall}a, shows the average of residual autocorrelation values across
$10^5$ independent realizations of $Y$ from a suite of simulations at each model order 
$M < 50$.  There is structure.

As well as being useful for the efficient calculation of $\rho_{rr}(l)$, the sequence
$(\tilde{R} \, \tilde{R}^*)$ is a powerful equivalent representation.
It defines the power spectral density of our sequence of residuals, which we denote
\begin{equation}
P_{rr}(k) = (\tilde{R} \, \tilde{R}^*)_k.
\end{equation}
For later convenience we will also define a corresponding \emph{correlation} spectral power
density $\tilde{\rho}_{rr}(k) \triangleq P_{rr}(k)
/ \left( \sum_{n=0}^{N-1} r^2_n \right)$, the DFT of $\rho_{rr}(l)$.  $P_{rr}(k)$ and
$\tilde{\rho}_{rr}(k)$ are both even symmetric functions of $|k|$ for purely real residuals.
For i.i.d.\ residuals with a defined variance, these functions are expected to show a flat,
\emph{white noise} spectrum with constant expectation values
$E(P_{rr}) = \sum_{n=0}^{N-1} r^2_n,
E(\tilde{\rho}_{rr}) = 1$ for all $k=0,\ldots, N-1$.

Figure~\ref{fig:scorall}b shows the averaged power spectral signature of residuals
from the same suite of simulations as shown in Figure~\ref{fig:scorall}a, where we have defined
the power spectral signature as
\begin{equation}
S_{rr}(k) = P_{rr} (k) / \max_k{\left\{P_{rr}\right\}},
\end{equation}
for some sequence of residuals $R$ with power spectral density $P_{rr}(k)$.  This choice makes the
structure of the power spectrum easy to visualize: with increasing $M$, we see that an increasing
share of the lower-$k$ spectral modes are suppressed, starting with the lowest $k$.  The apparent
structure in the modelling residuals from overfitting arises because
these residuals no longer match the flat power spectrum distribution that would be expected if they
could be assumed to be i.i.d.\ -- in a very real sense they have been \emph{high-pass filtered}.

\subsection{A Chebyshev Polynomial Series Model}\label{sec:chebysims}
The Fourier Series model \eqref{eq:sinusoids} is a special case:
it forms an othonormal basis, its
coefficients $a_m$ and $b_m$ have direct correspondence to
the real and imaginary parts of the DFT itself, and it provides an optimally
compact (in terms of $M$) series representation of arbitrary real-valued
sequences $R$.

The suppression of lower-$k$ power in
$P_{rr}(k)$ by overfitting models in the simulations performed above is therefore,
in fact, effectively \emph{total}; not only on average but also in individual cases.
In general models will often not have
complete freedom to precisely replicate arbitrary variation in observed data
(although see \citep{zhangetal17}), nor will they so wholly suppress Fourier modes in
residuals.

For this reason it is interesting to generalize to a set of model basis functions
that do not match the modes of the DFT precisely, and cannot represent arbitrary
patterns of residuals in any finite series.
Candidates for such a basis are the Chebyshev polynomials of the
first kind, $T_m(x)$ (see \citep{arfkenetal13}).
We define the Chebyshev polynomial series model of order $M$ as:
\begin{equation}\label{eq:chebyshev}
f_{\mathrm{c}, M} \left(x | a_0, a_1, \ldots, a_{M-1} \right) =
\sum_{m=0}^{M-1} a_m T_m(x).
\end{equation}
The absolute values of these functions are bounded by 1 on the interval $[-1, 1]$, and they are a 
convenient basis for polynomial interpolation in bounded spaces.
As for the Fourier Series, we simulate a simple case of $N=100$ observations in our sample $Y$.
We place these at corresponding evenly-spaced locations $x_n = -1 + 2n/N$
for $n=0,\ldots,{N-1}$.

\begin{figure}
  \centering
  \includegraphics[width=0.49\textwidth]{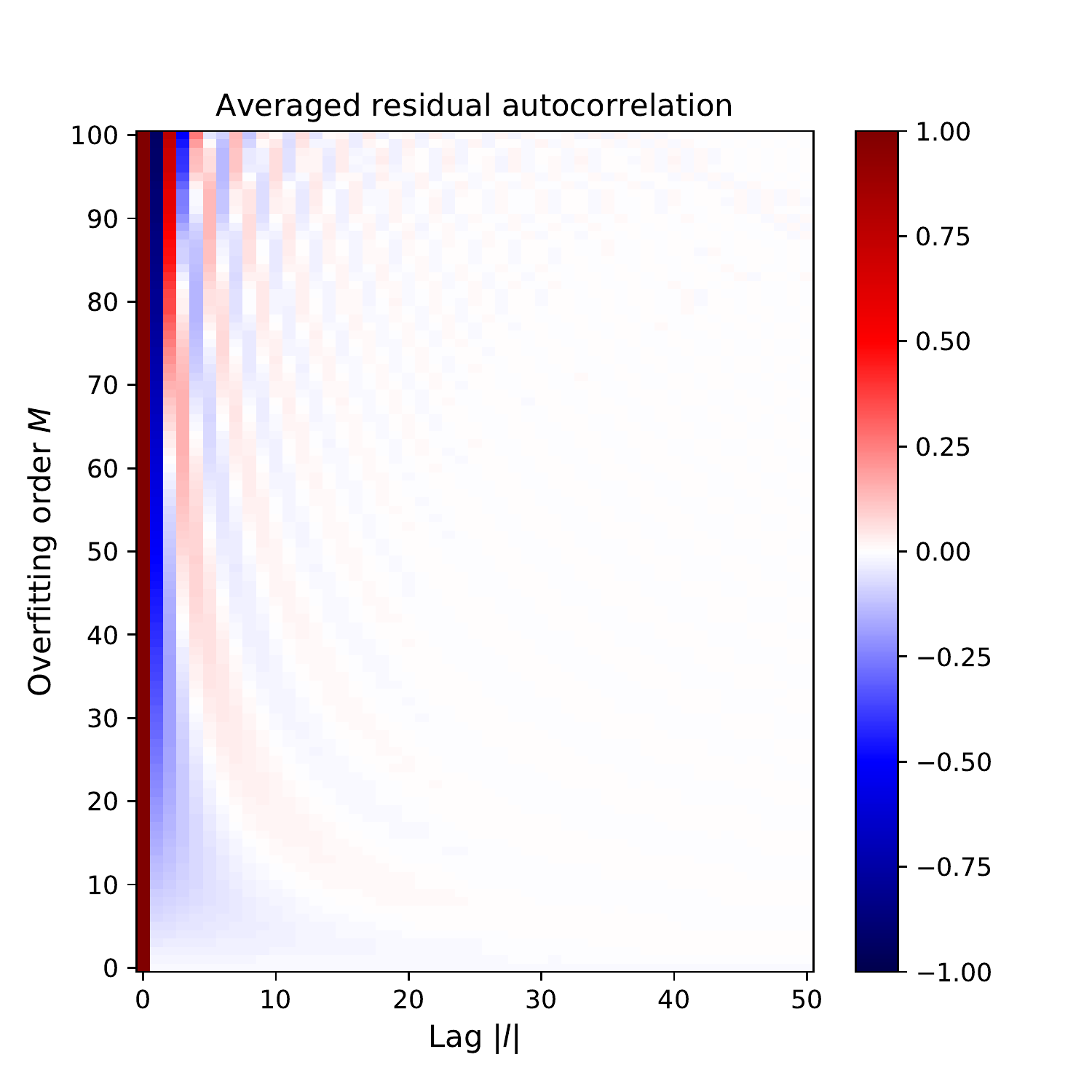}
  \includegraphics[width=0.49\textwidth]{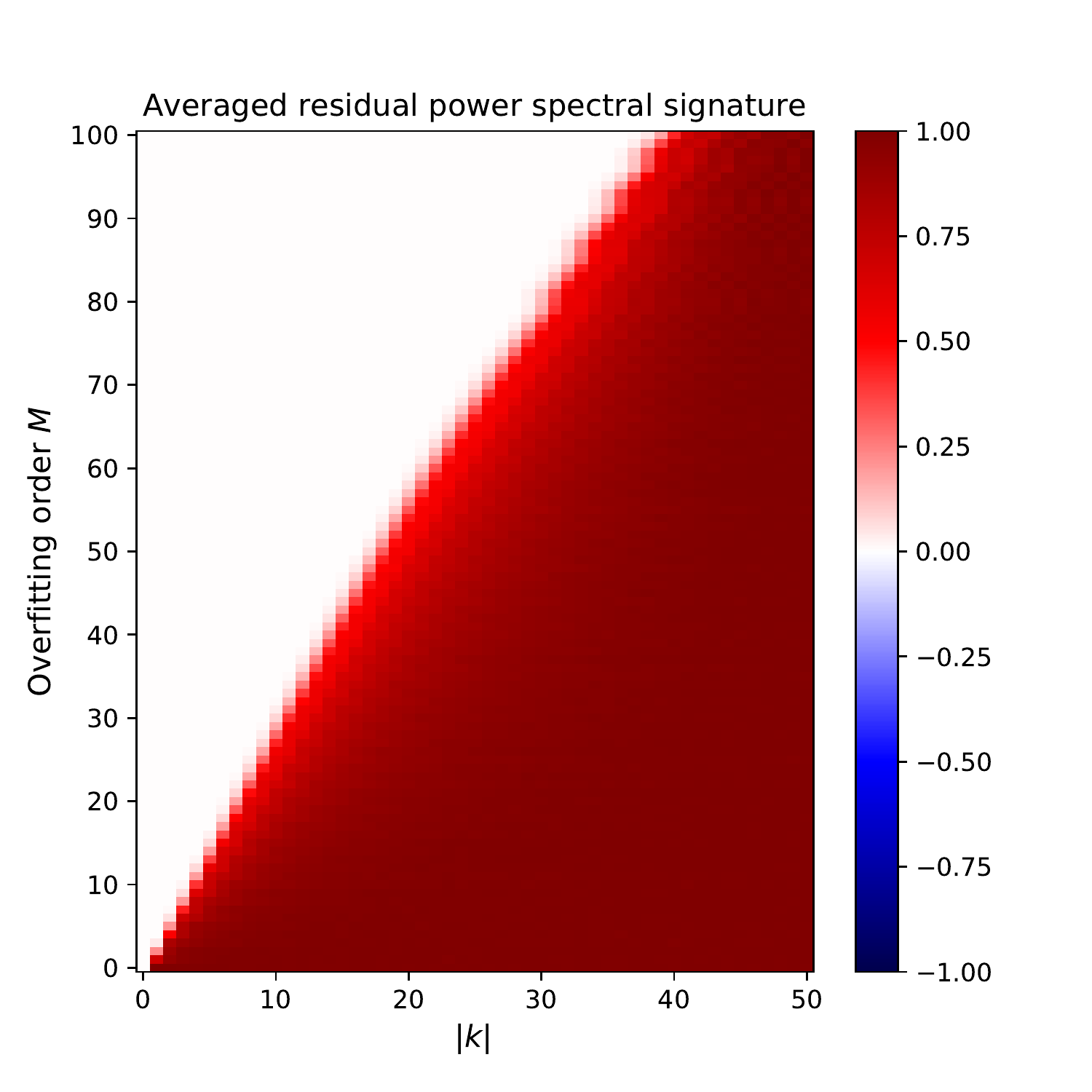}
  \caption{a) Averaged residual autocorrelation function values as a function of Chebyshev
  polynomial model order $M$,
  calculated over $10^5$ realizations of $Y$ per order; b) Averaged residual power spectral
  signature for the same realizations and model.\label{fig:ccorall}}
\end{figure}

Once more, we performed $10^5$ independent realizations of $Y$ for each Chebyshev model order
$M \leq 100$ (as the Chebyshev polynomial series model does not afford a complete representation
of an arbitrary sequence $R$ at any level of truncation, unlike the Fourier Series, the choice of
maxmimum $M$ is somewhat arbritrary).
Figure~\ref{fig:ccorall} shows the averaged residual autocorrelation $\rho_{rr}(l)$ and power
spectral signature $S_{rr}(k)$, as a function of order $M$, across this suite of simulations.
We see similar behaviour to the Fourier Series case, with increasing modelling order $M$ leading to
increasing supression of lower-$k$ power.  Due to the less compact representation of arbitrary
sequences provided by Chebyshev polynomials, the suppression of lower-$k$ power by overfitting is
more gradual.

\section{Implications for the Mean Squared Error Loss Function}
As discussed in Section~\ref{sec:intro}, the MSE loss function is very commonly used in
optimization and regression applications, and is defined using the notation above as simply
$\textrm{MSE}(R) = \frac{1}{N} \sum_{n=0}^{N-1} r^2_n$.
Underpinning the widespread use of the MSE as a loss function is the
Gauss-Markov theorem: that for a linear model fit to observations with uncorrelated, equal variance
and zero mean errors, the best linear unbiased estimator of any linear combination of the
observations is that which minimizes the MSE.
Where observational errors can be assumed to also follow a Gaussian distribution, this estimator is
also the Maximum Likelihood Estimator (e.g.\ \citep{pressetal92}).

But we have seen that residuals are not necessarily uncorrelated.  The model residuals passed to the
MSE loss function will only be uncorrelated in general if the `true' model has been applied, and
the model parameters $\bm{\theta}$ are optimal.
Anything else will, in general, either underfit or overfit, introducing correlations
as shown above and as discussed (in a specific application) by \citep{rowe10}. 
In empirical settings and
machine learning applications the `true' model may be unknown and potentially unknowable.  The
application of the MSE loss function in regression tasks, often by default, assumes too
much in assuming that model residuals are uncorrelated -- irrespective of model selection or
under variation in model parameter values -- however plausible this assumption may be
about the underlying observational errors.  This causes us to unwittingly discard valuable
information.

The information discarded is not captured by the MSE, which is invariant under
permutation of any ordered sequence of the residual, because it is embedded within
knowledge of the relative \emph{locations} of residuals with respect to one another.
The aim of this paper is to make initial step towards using this information in the optimization
process so as to build better regularized, and better generalizing, models.  If this can be done at
all, computational efficiency and ease of adoption will be valuable characteristics of the solution.
With these aims in mind,
we will explicitly target as simple a modification of the widely-used MSE loss function as possible.

\section{Incorporating Knowledge of Correlated Residuals into the Loss Function}
\subsection{A Multivariate Gaussian Distribution \& The Least Squares Case}
We begin with the simplest possible applicable probabily distribution for correlated residuals,
a multivariate
Gaussian.  We write the probability density function as
\begin{equation}\label{eq:pr}
p(r_0, \dots, r_{N-1}) = \frac{1}{\sqrt{(2 \pi)^N \det{(\mathbf{\Sigma})}}}
    \exp{
    \left[ - \frac{1}{2} \left( \mathbf{r} - \bm{\mu} \right)^{\mathrm{T}} \mathbf{\Sigma}^{-1}
    \left( \mathbf{r} - \bm{\mu} \right) \right]}
\end{equation}
where for convenience we have adopted vector notation to describe the sequence of residuals
$\mathbf{r} = (r_0, \ldots, r_{N-1})^{\mathrm{T}}$, with corresponding mean $\bm{\mu}$ and
covariance matrix $\mathbf{\Sigma}$.  We will assume $\bm{\mu} = 0$.

For ordinary least squares it would be additionally assumed that
$\mathbf{\Sigma} = \sigma^2 \mathbf{I}_N$, where $\mathbf{I}_N$ is the $N \times N$ identity matrix. 
and $\sigma^2$ is the variance of errors.  Then identifying \eqref{eq:pr} with the statistical
likelihood of the model
$\mathcal{L}(\mathbf{y}, \hat{\mathbf{y}} | \bm{\theta}, \sigma^2)$
leads to the log-likelihood function
\begin{equation}\label{eq:logl}
\ln{\left[ \mathcal{L}(\mathbf{y}, \hat{\mathbf{y}} | \bm{\theta}, \sigma^2 ) \right]} = 
-\frac{N}{2 \sigma^2} \textrm{MSE}(\mathbf{y}, \hat{\mathbf{y}} | \bm{\theta}) -
\frac{N}{2} \ln{ \left( 2 \pi \sigma^2 \right)}.
\end{equation}
Minimizing the MSE therefore maximizes the likelihood, irrespective of any value taken by the
(likely unknown) variance $\sigma^2$.

\subsection{A New Loss Function}
We will seek a new loss function of the form
$L(\mathbf{y}, \hat{\mathbf{y}} | \bm{\theta}) =
\textrm{MSE}(\mathbf{y}, \hat{\mathbf{y}} | \bm{\theta}) +
X(\mathbf{y}, \hat{\mathbf{y}} | \bm{\theta})$, where the
function $X$ is to be determined.  To identify a candidate, we will relax assumptions about the
covariance matrix of residuals in \eqref{eq:pr} relative to the least squares case,
and instead assume the slightly more general
\begin{equation}\label{eq:sigma2P}
\mathbf{\Sigma} = \sigma^2 \mathbf{P}
\end{equation}
where $\mathbf{P}$ is an $N \times N$, symmetric, positive
semidefinite correlation matrix with unit values on the diagonals.
We will have little a priori knowledge of the contents of $\mathbf{P}$, which as we know from
Section~\ref{sec:sims} can vary with model and model parameters, as well as with any correlations
that may existing in the underlying observational errors.

We propose to weaken the common assumption that $\mathbf{P} = \mathbf{I}_n$ by instead introducing
the following form for the prior probability on $\mathbf{P}$:
\begin{equation}\label{eq:prior}
p(\mathbf{P}) \propto \exp{\left[\eta \times h(\mathbf{P})\right]},
\end{equation}
where $h(\mathbf{P}) = \frac{N}{2}(1 + \ln{2\pi}) + \frac{1}{2} \ln{[\det(\mathbf{P})]}$
is the differential entropy for a multivariate Gaussian with covariance matrix
$\mathbf{P}$, and $\eta$ is a scaling constant that we introduce to allow the user to balance the
contribution to the loss function from this prior.  The prior can then be simplified to
\begin{equation}
p(\mathbf{P}) \propto \left[ \det{( \mathbf{P})} \right]^{\frac{\eta}{2}}.
\end{equation}
We motivate the functional form of \eqref{eq:prior}
by direct analogy to the Boltzmann entropy formula (e.g.\ \citep{hill12}), in the intuitive
belief that the most likely distribution of residuals is that which has the greatest number of
possible configurations per unit of variance.
As $\det{(\mathbf{P})}$ is maximized when $\mathbf{P} = \mathbf{I}_n$, applying this prior
will penalize distributions of residuals showing any correlations, whether from overfitting
or underfitting \citep{rowe10}.

Crucially, we will also assume that $\mathbf{P}$ can be approximated as a circulant matrix.
(Although not true in general, particularly for small samples, the approximation becomes
asymptotically more accurate as $N$ increases and/or as residual correlations themselves become
negligible, since $\mathbf{I}_N$ is a circulant matrix.)
Any circulant matrix $\mathbf{P}$ is fully determined by its first column which we label
$\bm{\rho}$.

As $\mathbf{P}$ is a correlation matrix, the column $\bm{\rho}$ consists of the
`true' values around which the
(circular) residual autocorrelation function $\rho_{rr}(l)$ of
equation~\eqref{eq:rho}, defined for a single sample of residual values, will be distributed.
It can be seen that the average of $\rho_{rr}(l)$ over realizations of residuals
from multiple independent samples of observations $Y$, given the same model, will form an
asymtotically consistent estimator for $\bm{\rho}$.
But for any sufficiently free, circulant matrix model of $\mathbf{P}$ even a single sample of
residuals allows us to calculate a $\rho_{rr}(l)$ that can itself be identified as the maximum
likelihood estimate of $\bm{\rho}$; we will denote this estimate $\bar{\bm{\rho}} = \rho_{rr}(l)$.
If $\mathbf{P}$ is assumed circulant this also defines a corresponding maximum likelihood
estimate for the full matrix, $\bar{\mathbf{P}}(\mathbf{y}, \hat{\mathbf{y}} | \bm{\theta})$.

If $\mathbf{P}$ is a circulant matrix its eigenvalues are the
DFT of $\bm{\rho}$.  The same holds for the eigenvalues of $\bar{\mathbf{P}}$:
\begin{equation}
(\lambda_{\bar{\mathbf{P}}})_k =  \left[ \mathcal{F} \left\{ \bar{\bm{\rho}} \right\} \right]_k =
\tilde{\rho}_{rr}(k),
\end{equation}
where $k=0,\ldots,N-1$.  Since the determinant of a matrix is given by the product of its
eigenvalues, $\det{(\bar{\mathbf{P}})}$ can be
written as
\begin{equation}
\det{\left[ \bar{\mathbf{P}}(\mathbf{y}, \hat{\mathbf{y}} | \bm{\theta}) \right]} =
\prod_{k=0}^{N-1} (\lambda_{\bar{\mathbf{P}}})_k
= \prod_{k=0}^{N-1} \tilde{\rho}_{rr}(k).
\end{equation}

As discussed above, the MSE loss function can be related to the log-likelihood
function of the
multivariate Gaussian with uncorrelated residuals.  In a full, formally-correct treatment this
likelihood should be expanded to include a freely-varying $\mathbf{P}$ which can then be
marginalized over as a nuisance parameter, combining both the
prior $p(\mathbf{P})$ and the likelihood of observed correlated residuals \emph{given}
$\mathbf{P}$ in the integrand.  But this is reserved for future work.
Instead we propose an approximate treatment that assumes that the likelihood of any
observed residual correlations given some $\mathbf{P}$ can be approximated as very sharply peaked,
indeed as being a Dirac
delta function $\delta\left(\mathbf{P} - \bar{\mathbf{P}}\right)$, around $\bar{\mathbf{P}}$.
This makes the marginalization trivial to compute, and adding a penalty term
$-\ln{\left[p(\bar{\mathbf{P}})\right]}$
to the multivariate Gaussian log-likelihood of \eqref{eq:logl} to create new loss function:
\begin{equation}
L(\mathbf{y}, \hat{\mathbf{y}} | \bm{\theta}) = 
-\ln{\left[ \mathcal{L}(\mathbf{y}, \hat{\mathbf{y}}|\bm{\theta}) \right]}
-\ln{\left[p(\bar{\mathbf{P}}(\mathbf{y}, \hat{\mathbf{y}} | \bm{\theta}))\right]} + \ldots ,
\end{equation}
where we will now omit terms that have no dependence on the parameter vector $\bm{\theta}$.
Dividing both sides by $N/\sigma^2$ we arrive at the equivalent
\begin{equation}
L(\mathbf{y}, \hat{\mathbf{y}} | \bm{\theta}) =
 \textrm{MSE}(\mathbf{y}, \hat{\mathbf{y}} | \bm{\theta}) - \sigma^2 \frac{\eta}{N}
 \sum_{k=0}^{N-1}  \ln{\left[ \tilde{\rho}_{rr}(k) \right]}.
\end{equation}
Defining the Mean Log Power of residual correlations as
$\textrm{MLP} = \frac{1}{N}\sum_{k=0}^{N-1} \ln{ \left[\tilde{\rho}_{rr}(k)\right]}$,
and making a final
approximation that the MSE is a consistent, sufficient estimator for $\sigma^2$
(valid in the limit $\mathbf{P} \rightarrow \mathbf{I}_n$) we arrive at
the following proposal for a residual entropy-sensitive extension of the MSE loss function:
\begin{equation}\label{eq:loss}
L(\mathbf{y}, \hat{\mathbf{y}} | \bm{\theta}) =
\textrm{MSE}(\mathbf{y}, \hat{\mathbf{y}} | \bm{\theta}) \times
\left[ 1 - \eta \: \textrm{MLP}(\mathbf{y}, \hat{\mathbf{y}} | \bm{\theta}) \right].
\end{equation}

\begin{figure}
  \centering
  \includegraphics[width=0.49\textwidth]{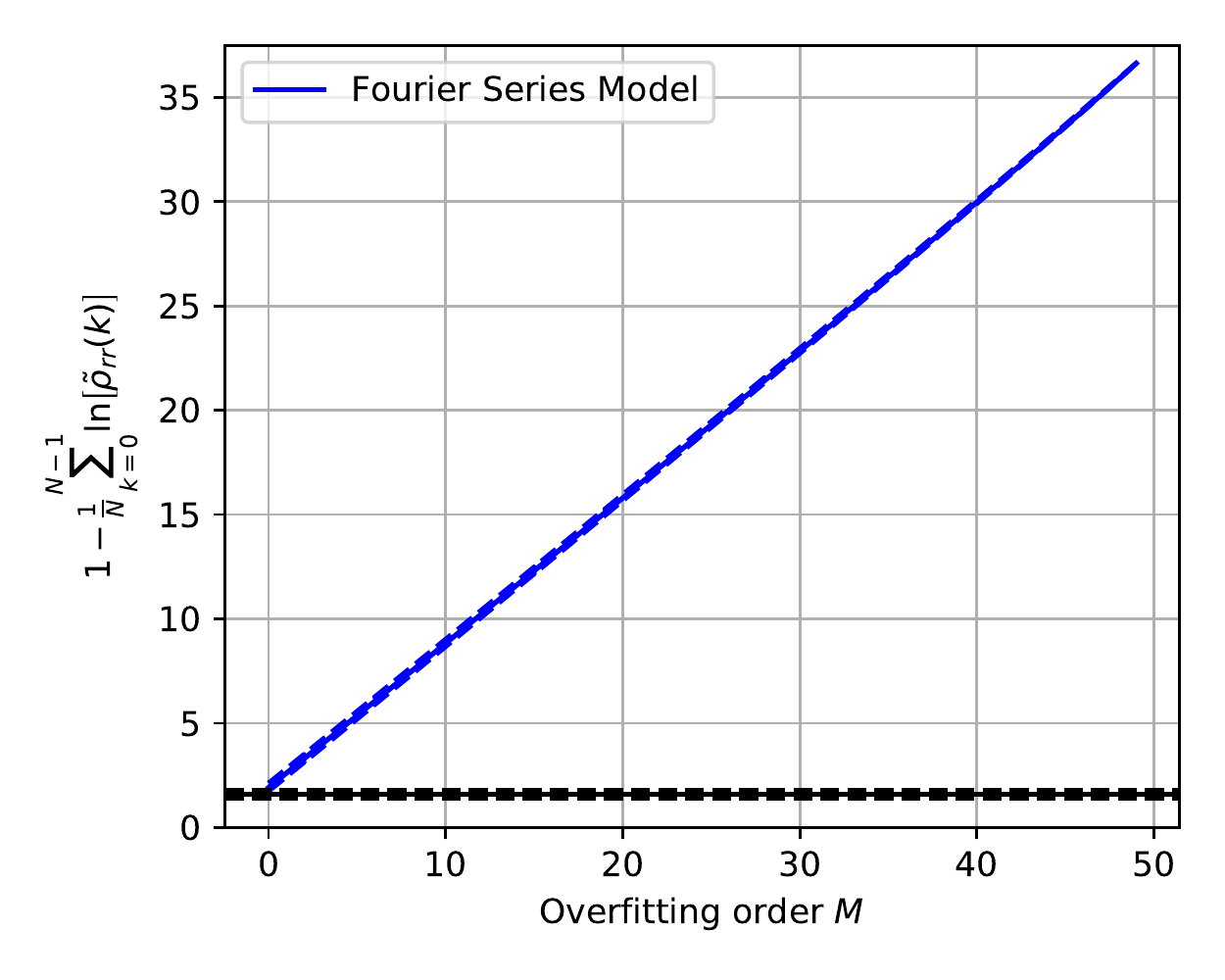}
  \includegraphics[width=0.49\textwidth]{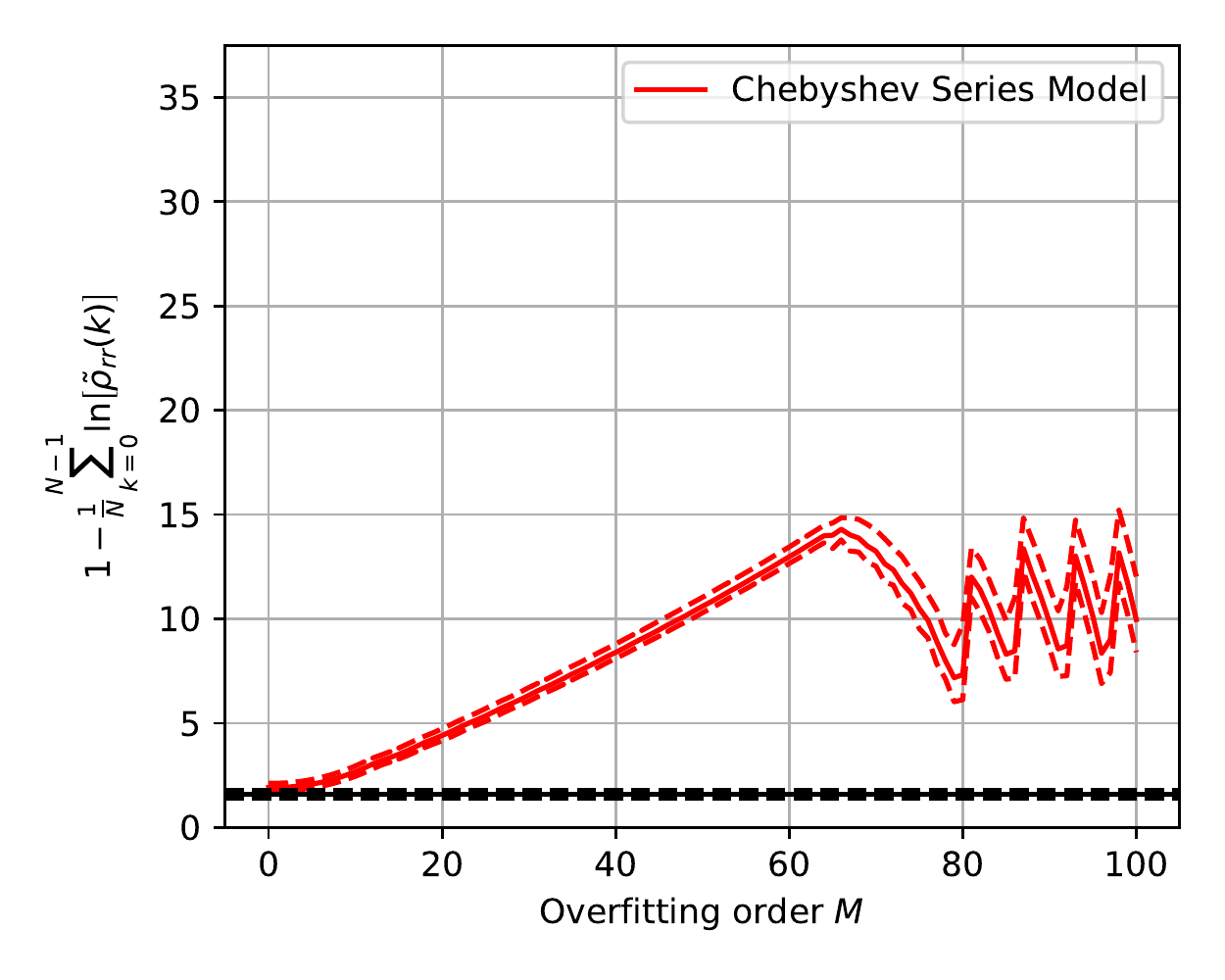}
  \caption{a) Median (solid blue line) and 5th, 95th percentile boundaries (dashed blue lines)
  of values of $\left[1 - \textrm{MLP}(R) \right]$ for the simulations of
  Section~\ref{sec:sinsims}, as a function of
  Fourier Series model order $M$: the black lines show the distribution of this quantity estimated
  from $Y$ in the absence of any model fit; b)
  The same but for the Chebyshev simulations of Section~\ref{sec:chebysims} (red lines):
  following
  investigation we have concluded that the structure seen at high $M$ values is due to low-level
  aliasing artifacts introduced by the significant variation in the higher-order Chebyshev
  polynomials at spatial frequencies higher than the Nyquist frequency.
  These artifacts are also present (but not visible) in the low-$k$ modes of
  Figure~\ref{fig:ccorall}.\label{fig:mlp}}
\end{figure}
Figure~\ref{fig:mlp} shows statistics of realized values
of $\left[1 - \textrm{MLP}(R)\right]$, i.e.\ the proposed multiplier on the MSE in the
$\eta = 1$ case,
for every residual sample $R$ in the simulations of Section~\ref{sec:sims}.
We see an increasing function of overfitting order $M$ (although unavoidable
aliasing in the experimental setup introduces artifacts at
very high $M$ in the Chebyshev case), motivating the use of our proposed loss function
\eqref{eq:loss} in the suppression of residual correlations, and thereby
providing automatic regularization and a safeguard against overfitting.

\section{Conclusions}
Least squares has been the dominant approach towards empirical
regression problems for over 200 years \citep{legendre1805,pressetal92}, and the MSE loss function
is at its core.  
Yet we do believe that the MSE discards
valuable information about the locations of residuals.  We also believe that penalizing models
that show discernible structure in residuals, residuals that have hitherto been blindly assumed to
be uncorrelated, is both intuitively appealing and can be theoretically justified on entropic
principles. 

We have made a first attempt towards incorporating these beliefs directly into the
loss function.  We presented results that suggest the MSE may be improved by a simple (although
approximate) multiplicative factor given in equation \eqref{eq:loss}, a factor that
can be calculated with relative computational efficiency thanks to the FFT algorithm.  But
many issues and outstanding questions remain.

We have made mathematical approximations that are asymptotically true,
e.g.\ $\mathbf{P}$ being approximately circulant (valid only for large
$N$ or as $\mathbf{P} \rightarrow \mathbf{I}_n$) or that the MSE forms
a consistent and sufficient estimator for $\sigma^2$ in practice (valid as
$\mathbf{P} \rightarrow \mathbf{I}_n$).
Assumptions were also made to avoid marginalization over $\mathbf{P}$ and to
retain the MSE as part of the proposed loss function,
approximating the likelihood function of residual correlations as being
sharply distributed around the (observed) maximum likelihood estimate.
The impact of these simplifying assumptions on practical optimization needs to
be tested, and a fuller treatment may in fact lead to a better approximate form
by which to encourage greater residual entropy.

To calculate a meaningful power spectral density of residuals, the residuals must first be ordered
into a sequence that has some meaning in the input space.  This is trivial in one dimension,
less so in higher dimensions.  For higher dimensional regression problems we would initially
propose using residuals ordered along each of the most significant principal components of the
input space, in turn, as determined via PCA. 
Structure seen in residuals along any of the principal directions
in the space of input observations should be penalized, but whether this will be a powerful
enough constraint is an open question.  For
very high dimensional problems, it might be prudent to work in configuration (lag) space,
and estmate correlations in annular bins of cosine distance.
Where data are not regularly spaced in the independent
variables it may also be valuable to average binned residuals before calculating
$\tilde{\rho}_{rr}(k)$.

Clearly, an important next step will be applying the loss function of equation \eqref{eq:loss} in
practical settings, to explore its behaviour outside of artificial simulations and the
dynamic interplay of the MSE and the multiplicative factor
$\left[ 1 - \eta \: \textrm{MLP}\right]$.  Of utmost importance will be to establish whether the
new addition to the loss function is `safe', i.e.\
that modelling outcomes are no worse than when using MSE alone.

Finally, an important application not explored here is that of outlier rejection -- to the extent
that contaminating outliers drag a best-fitting model away from the bulk of the other data points,
they will necessarily introduce (positive) correlations in modelling residuals. 
The loss function described above offers hope of systematically enhancing robustness to outliers by
penalizing such fits automatically.

%

\bibliographystyle{plain}
\small
\bibliography{resent_neurips_2019}

\begin{thebibliography}{10}

\bibitem{arfkenetal13}
G.~B. Arfken, H.~J. Weber, and F.~E. Harris.
\newblock {\em Mathematical Methods for Physicists: A Comprehensive Guide}.
\newblock Elsevier Science, 2013.

\bibitem{berger85}
J.~O. Berger.
\newblock {\em {Statistical decision theory and Bayesian analysis; 2nd ed.}}
\newblock Springer Series in Statistics. Springer, New York, 1985.

\bibitem{boxpierce70}
G.~E.~P. Box and D.~A. Pierce.
\newblock Distribution of residual autocorrelations in
  autoregressive-integrated moving average time series models.
\newblock {\em Journal of the American Statistical Association},
  65(332):1509--1526, 1970.

\bibitem{breusch78}
T.~S. Breusch.
\newblock Testing for autocorrelation in dynamic linear models.
\newblock {\em Australian Economic Papers}, 17(31):334--355, 1978.

\bibitem{cooleytukey65}
J.~Cooley and J.~Tukey.
\newblock An algorithm for the machine calculation of complex fourier series.
\newblock {\em Mathematics of Computation}, 19(90):297--301, 1965.

\bibitem{durbinwatson50}
J.~Durbin and G.~S. Watson.
\newblock Testing for serial correlation in least squares regression. i.
\newblock {\em Biometrika}, 37(3-4):409--428, 12 1950.

\bibitem{durbinwatson51}
J.~Durbin and G.~S. Watson.
\newblock Testing for serial correlation in least squares regression. ii.
\newblock {\em Biometrika}, 38(1-2):159--178, 06 1951.

\bibitem{durbinwatson71}
J.~Durbin and G.~S. Watson.
\newblock Testing for serial correlation in least squares regression. iii.
\newblock {\em Biometrika}, 58(1):1--19, 1971.

\bibitem{godfrey78}
L.~G. Godfrey.
\newblock Testing against general autoregressive and moving average error
  models when the regressors include lagged dependent variables.
\newblock {\em Econometrica}, 46(6):1293--1301, 1978.

\bibitem{godfreytremayne05}
L.~G. Godfrey and A.~R. Tremayne.
\newblock The wild bootstrap and heteroskedasticity-robust tests for serial
  correlation in dynamic regression models.
\newblock {\em Computational Statistics \& Data Analysis}, 49(2):377--395, 4
  2005.

\bibitem{hill12}
T.~L. Hill.
\newblock {\em An Introduction to Statistical Thermodynamics}.
\newblock Dover Books on Physics. Dover Publications, 2012.

\bibitem{legendre1805}
A.~M. Legendre.
\newblock {\em Nouvelles m{\'e}thodes pour la d{\'e}termination des orbites des
  com{\`e}tes}.
\newblock Nineteenth Century Collections Online (NCCO): Science, Technology,
  and Medicine: 1780-1925. F. Didot, 1805.

\bibitem{ljungbox78}
G.~M. Ljung and G.~E.~P. Box.
\newblock On a measure of lack of fit in time series models.
\newblock {\em Biometrika}, 65(2):297--303, 08 1978.

\bibitem{pressetal92}
W.~H. {Press}, S.~A. {Teukolsky}, W.~T. {Vetterling}, and B.~P. {Flannery}.
\newblock {\em {Numerical recipes in FORTRAN. The art of scientific
  computing}}.
\newblock 1992.

\bibitem{proia18}
F.~Proïa.
\newblock Testing for residual correlation of any order in the autoregressive
  process.
\newblock {\em Communications in Statistics - Theory and Methods},
  47(3):628--654, 2018.

\bibitem{rowe10}
B.~{Rowe}.
\newblock {Improving PSF modelling for weak gravitational lensing using new
  methods in model selection}.
\newblock {\em Mon.\ Not.\ Roy.\ Astron.\ Soc.}, 404:350--366, May 2010.

\bibitem{vonneumann41}
J.~von Neumann.
\newblock Distribution of the ratio of the mean square successive difference to
  the variance.
\newblock {\em Ann. Math. Statist.}, 12(4):367--395, 12 1941.

\bibitem{zhangetal17}
C.~Zhang, S.~Bengio, M.~Hardt, B.~Recht, and O.~Vinyals.
\newblock Understanding deep learning requires rethinking generalization.
\newblock {\em CoRR}, abs/1611.03530, 2017.

\end{thebibliography}

\end{document}